\documentclass[preprint,aps,showpacs,prb]{revtex4}
 
\usepackage{graphicx}
\usepackage{dcolumn}
\usepackage{bm}
 
\begin{document}

\preprint{Submitted to PRB}
\title{Density-matrix functional theory of strongly-correlated lattice 
fermions}
\author{R.~L\'opez-Sandoval}
\affiliation{Instituto Potosino de Investigaci\'on Cient\'{\i}fica 
y Tecnol\'ogica, 
Av.\ Venustiano Carranza 2425-A, 78210 San Luis Potos\'{\i}, M\'exico}
\author{G.~M.~Pastor}
\affiliation{Laboratoire de Physique Quantique, Universit\'e Paul Sabatier, 
Centre National de la Recherche Scientifique, 118 route de Narbonne,
31062 Toulouse, France}

\date{\today}

\begin{abstract}
A density functional theory (DFT) of lattice fermion models is 
presented, which uses the single-particle density matrix $\gamma_{ij}$ 
as basic variable. A simple, explicit approximation to the 
interaction-energy functional $W[\gamma]$ of the Hubbard model
is derived from exact dimer results, scaling 
properties of $W[\gamma]$ and known limits. Systematic tests on the
one-dimensional chain show a remarkable agreement with the 
Bethe-Ansatz exact solution for all interaction regimes and band fillings.
New results are obtained for the ground-state energy
and charge-excitation gap in two dimensions. 
A successful description of strong electron correlations within 
DFT is achieved.
\end{abstract}
%



%
\maketitle

\section{\label{sec:introd}Introduction}

First principles methods and many-body lattice models are the two
main theoretical approaches to the electronic properties of matter. 
From the first-principles perspective, the major breakthrough
in the last decades has been Hohenberg-Kohn-Sham's (HKS) 
density-functional theory (DFT) and the derived powerful 
methods of electronic-structure calculation.\cite{parr-book}
Despite their unparalleled success in an extremely wide variety 
of problems, current implementations of DFT have still serious 
difficulties in accounting for phenomena that involve strong electron 
correlations as observed, for example, in heavy-fermion materials, 
Mott insulators or high-$T_c$ superconductors.\cite{mb-books}
Being in principle an exact theory, the limitations of DFT  
have to be ascribed to the approximations used for the 
interaction-energy functional $W[\rho(\vec{r})]$
and not to the underlying formalism. The development of
new functionals improving  the description of strong correlation 
effects is therefore a major current theoretical challenge.

On the other side, the physics of strongly-correlated Fermi systems 
is intensively studied in the framework of parametrized lattice models 
(e.g., Hubbard, Anderson, etc.) by using specific leading-edge many-body 
techniques.\cite{mb-books} Taking into account the universality of DFT,
and its demonstrated efficiency in complex {\em ab initio} 
calculations, it is quite remarkable that only  few investigations
have been concerned so far with applying the concepts 
of DFT to the lattice models describing strongly correlated 
fermions.\cite{gunn,godby,carl,prbxcfun} In fact, already from a 
formal standpoint, one may expect that DFT with an appropriate Ansatz 
for $W$ should be a particularly valuable many-body approach to 
lattice models, thus becoming a subject of theoretical interest on its own.
Moreover, DFT studies on simpler universal models also 
provide useful new insights relevant to first principles 
calculations,\cite{parr-book} particularly since in some cases
the exact solution of the many-body problem is available.\cite{mb-books}

The purpose of this paper is to extend the scope of DFT to the description 
of strong electron correlations in lattice Hamiltonians and to demonstrate 
quantitatively for the first time the performance of lattice 
density-functional theory (LDFT) in one-dimensional (1D) and 
two-dimensional (2D) systems. 
Sec.~\ref{sec:LDFT} presents concisely the basic formalism of LDFT. 
In this framework the ground-state properties 
are obtained from the solution of exact self-consistent equations 
that involve derivatives of the interaction-energy functional 
$W[\gamma]$ with respect to the single-particle density matrix $\gamma$.
In Sec.~\ref{sec:Wdim} the dependence of W on the 
nearest-neighbor (NN) density-matrix element $\gamma_{12}$ is analyzed 
and a simple explicit approximation to $W(\gamma_{12})$ is derived 
for the Hubbard model. Sec.~\ref{sec:res} discusses representative 
applications of this {\em Ansatz}. First, the accuracy of the method is 
demonstrated by comparison with available exact results on the 1D 
Hubbard model. New results are then discussed, particularly concerning the 
ground-state energy and charge-excitation gap in 2D lattices. 
Finally, Sec.~\ref{sec:concl} summarizes our conclusions.

\section{\label{sec:LDFT}Lattice density-functional theory}

In order to be explicit we focus on the Hubbard model which is expected 
to capture the main physics of lattice fermions in a narrow energy 
band. The Hamiltonian
\begin{equation}
\label{eq:hamhub}
H = \sum_{\langle i,j\rangle \sigma} t_{ij}
\hat c^{\dagger}_{i \sigma} \hat c_{j \sigma} +
U \sum_i  \hat n_{i \downarrow} \hat n_{i\uparrow} ,
\end{equation}
includes nearest neighbor (NN) hoppings $t_{ij}$, and an on-site 
interactions $U$  ($\hat n_{i\sigma} = 
\hat c_{i\sigma}^\dagger \hat c_{i\sigma}$).  
The importance of electron correlations is controlled by one 
parameter, namely, the ratio $U/t$.
The hopping integrals $t_{ij}$ are defined by the lattice structure
(typically, $t_{ij} = - t < 0$ for NN $ij$) and thus play the role given 
in conventional DFT to the external potential $V_{ext}(\vec r)$. 
Consequently, in LDFT 
the single-particle density matrix $\gamma_{ij}$ replaces the 
density $\rho(\vec r)$ as basic variable,  
since the hopping integrals $t_{ij}$ are nonlocal in the sites.\cite{godby} 
The situation is similar to the density-matrix functional
theory proposed by Gilbert for the study of nonlocal  
pseudopotentials.\cite{gilb,foot_local}

The ground-state energy $E_{gs}$ and density-matrix $\gamma_{ij}^{gs}$
are determined by minimizing the energy functional  
\begin{equation}
\label{eq:E}
E[\gamma] = E_K[\gamma] + W [\gamma]
\end{equation}
with respect to $\gamma_{ij}$.\cite{godby} The first term
\begin{equation}
\label{eq:EK}
E_K[\gamma] = \sum_{ij} t_{ij} \gamma_{ij}
\end{equation}
is the kinetic energy associated with the electronic motion in the 
lattice. The second term is Levy's interaction-energy functional\cite{levy}
given by
\begin{equation}
\label{eq:Wlevy}
W[\gamma] = 
\min \left[ U \sum_i  \langle \Psi [\gamma] |
\hat n_{i\uparrow} \hat n_{i\downarrow}|
\Psi [\gamma] \rangle  \right] \; ,
\end{equation}
where the minimization runs over all $N$-particle states 
$| \Psi [\gamma] \rangle$ satisfying
\begin{equation}
\label{eq:constr}
\langle \Psi [\gamma] | 
\sum_\sigma \hat c_{i \sigma }^{\dagger} \hat c_{j \sigma} 
|\Psi [\gamma] \rangle = \gamma_{ij}
\end{equation}
for all $i$ and $j$.\cite{godby,carl,prbxcfun,gilb} 
$W[\gamma]$ represents the minimum interaction energy compatible with a given 
$\gamma_{ij}$. It is a universal functional
in the sense that it is independent of $t_{ij}$. However, note that 
it still depends on the type of model interaction, 
on the number of electrons $N_e$ or band filling $n=N_e/N_a$, 
and on the number of sites $N_a$.\cite{foot_gen}

$E[\gamma]$ is minimized by expressing 
$\gamma_{ij} = \gamma_{ij\uparrow} + \gamma_{ij\downarrow}$ 
in terms of its eigenvalues $\eta_{k\sigma}$ (occupation numbers) 
and eigenvectors $u_{ik\sigma}$ (natural orbitals) as
\begin{equation}
\label{eq:gamma}
\gamma_{ij\sigma} = \sum_k u_{ik\sigma} \eta_{k\sigma} u_{jk\sigma}^* \; .
\end{equation}
Introducing Lagrange multipliers $\mu$ and $\lambda_{k\sigma}$ 
($\varepsilon_{k\sigma}=\lambda_{k\sigma}/\eta_{k\sigma}$) 
to impose the usual constraints 
$\sum_{k\sigma} \eta_{k\sigma} = N_e$ and 
$\sum_i\vert u_{ik\sigma}\vert^2 = 1$, one obtains 
the eigenvalue equations 
\begin{equation}
\label{eq:minsc}
\sum_j \left(t_{ij} + \frac{\partial W} {\partial \gamma_{ij \sigma}} \right) 
u_{jk\sigma} = \varepsilon_{k\sigma} u_{ik\sigma}
\end{equation}
with 
$\varepsilon_{k\sigma}  <  \mu $   ($\varepsilon_{k\sigma} > \mu $) 
if $\eta_{k\sigma} = 1$ ($\eta_{k\sigma} = 0$), and 
\begin{equation}
\label{eq:eps=mu}
\varepsilon_{k\sigma} = \mu \quad {\rm if} \quad  0<\eta_{k\sigma} < 1\; .
\end{equation}   
In Eq.~(\ref{eq:minsc}) self-consistency is implied by the dependence of 
$\partial W / \partial\gamma_{ij \sigma}$ on $\eta_{k\sigma}$ and 
$u_{ik\sigma}$. The present formulation is analogous to well-known 
results of density-matrix functional
theory in the continuum.\cite{gilb} However, notice the 
fundamental differences with respect to the KS-like approach proposed 
in Ref.~\cite{godby}, which assumes non-interacting 
$v$-representability, and where only integer occupations are allowed.
The importance of fractional orbital occupations to the description of 
electron correlations within density-matrix functional
theory has already been stressed by
Gilbert.\cite{gilb} In particular for the Hubbard model, one observes 
that $0 < \eta_{k \sigma} < 1$ for all $k$, except in very special 
situations such as  $U/t=0$ or the fully-polarized ferromagnetic state.
This can be understood from perturbation-theory arguments
---none of the $\eta_{k\sigma}$ is a 
good quantum number for $U/t\not= 0$--- and has been explicitly 
demonstrated in exact solutions for finite clusters or the 
1D chain.\cite{lieb-wu} Therefore, the case (\ref{eq:eps=mu})
is the only relevant one in general and all $\varepsilon_{k\sigma}$ 
in Eq.~(\ref{eq:minsc}) must be degenerate. Consequently, 
\begin{equation}
\label{eq:tij}
t_{ij} + \frac{\partial W} {\partial\gamma_{ij \sigma}} 
= \delta_{ij} \; \mu 
\end{equation} 
for all $i$ and $j$. Note that approximations of $W$ in terms of diagonal 
$\gamma_{ii}$ alone can never yield such a behavior.\cite{foot_local}

At this point it is important to observe that the general functional, 
valid for all lattice structures and for all types of hybridizations, 
can be simplified at the expense of universality if the hopping 
integrals are short ranged. For example, if only NN hoppings are 
considered, $E_K$ is independent of $\gamma_{ij}$ for pairs of sites $ij$ 
that are not NN's. In this case, the constraints 
$\langle \Psi [\gamma] | 
\sum_\sigma \hat c_{i \sigma }^{\dagger}\hat c_{j \sigma} 
|\Psi[\gamma]] \rangle = \gamma_{ij}$ 
in Eqs.~(\ref{eq:Wlevy}) and (\ref{eq:constr}) need to be imposed only 
for $i=j$ and for NN $ij$. This allows to reduce drastically the number 
of variables and simplifies considerably the search for practical 
approximations to $W$. 
Moreover, in periodic lattices the ground-state $\gamma_{ij}^{gs}$ is a 
translational invariant. In order to determine $E_{gs}$ and 
$\gamma_{ij}^{gs}$, one may then set $\gamma_{ii} = n = N_e/N_a$ for all 
sites $i$, and $\gamma_{ij} = \gamma_{12}$ for all NN pairs $ij$. 
Thus, the interaction energy can be regarded as a simple function 
$W(\gamma_{12})$ of the density-matrix element between NN's. It should 
be however noted that this also implies that $W$ loses its 
universal character, since the NN map and the resulting dependence 
of $W$ on $\gamma_{12}$ are in principle different for different 
lattice structures.\cite{prbxcfun}

\section{\label{sec:Wdim}Interaction-energy functional for the Hubbard model}

Given a self-consistent scheme that implements the variational principle,
the challenge is to find a good, explicit approximations to the 
interaction-energy functional. $W[\gamma]$ may be determined exactly 
for small clusters by using numerical methods that perform the 
constrained minimization explicitly.\cite{prbxcfun} 
For a Hubbard dimer with $N_e=N_a=2$ a straightforward analytical
calculations yields
\begin{equation}
\label{eq:W2ex}
\frac{W(\gamma_{12})}{N_a} = \frac{U} {4} 
\left( 1  - \sqrt{ 1 -  \gamma_{12}^2 } \right)   \; ,
\end{equation}
which represents the minimum average number of double occupations 
for a given degree of electron delocalization, i.e., for a given 
$\gamma_{12}$ ($U>0$). Despite its simplicity,  
Eq.~(\ref{eq:W2ex}) already includes the fundamental interplay 
between electron delocalization and charge fluctuations, and 
provides useful insights on several general properties of $W(\gamma_{12})$ 
that are valid for arbitrary lattices: \\ 
(i) The domain of definition of $W(\gamma_{12})$ is limited by the
pure-state representability of $\gamma_{12}$. In fact, 
$\gamma_{12} \le \gamma_{12}^0 = 1$, where $\gamma_{12}^0$ corresponds
to the extreme of the kinetic energy (maximum degree of delocalization)
and thus to the $U=0$ ground-state for a given lattice and a given $n$. \\ 
(ii) For $\gamma_{12} = \gamma_{12}^0$, the underlying electronic state
$\Psi[\gamma_{12}^0]$ is a single Slater determinant and therefore 
$W(\gamma_{12}^0)= E_{\rm HF} = n^2U/4$, where $E_{\rm HF}$ is the 
Hartree-Fock energy. Moreover, 
$\partial W / \partial \gamma_{12}=\infty$ for 
$\gamma_{12} = \gamma_{12}^0$, since $\gamma_{12}^{gs} < \gamma_{12}^0$
already for arbitrary small $U/t$, as expected from perturbation theory. \\
(iii) Starting from $\gamma_{12} = \gamma_{12}^0$, $W(\gamma_{12})$ 
decreases monotonically with decreasing $\gamma_{12}$ reaching 
its lowest possible value, $W=0$, for $\gamma_{12} = \gamma_{12}^\infty$ 
($\gamma_{12}^\infty = 0$ for $n=1$). The fact that $W$ 
decreases with decreasing $|\gamma_{12}|$ shows that the 
correlation-induced reduction of the Coulomb energy is obtained
at the expense of electron delocalization. \\
(iv) $\gamma_{12}^\infty$ represents the largest NN bond order that can 
be obtained under the constraint of vanishing Coulomb energy. A lower 
bound for $\gamma_{12}^{\infty}$ is given by the bond order 
$\gamma_{12}^{\rm FM}$ in the fully-polarized ferromagnetic state which 
is formed by occupying the $N_e$ lowest single-particle states of the 
same spin ($n\le 1$). Note that the ground-state
$\gamma_{12}^{gs}$ always satisfies 
$\gamma_{12}^\infty \le \gamma_{12}^{gs} \le \gamma_{12}^0$ even though,
for $n \not= 1$, it is possible to construct $N_e$-electron states having 
$|\gamma_{12}| < |\gamma_{12}^\infty|$.

In order to derive a simple approximation to $W(\gamma_{12})$
that preserves the previous general properties we take advantage of 
its scaling properties. Exact numerical studies\cite{prbxcfun} 
have shown that $W(\gamma_{12})$ depends weakly on $N_e$, $N_a$ and
lattice structure if it is measured in units of $E_{\rm HF}$ and if
$\gamma_{12}$ is scaled within the relevant domain of representability
$[\gamma_{12}^\infty,\gamma_{12}^0]$. Physically, this means that
the relative change in $W$ associated to a given change in the 
degree of electron localization 
$g_{12} = (\gamma_{12} - \gamma_{12}^\infty) / 
(\gamma_{12}^0 - \gamma_{12}^\infty)$ can be regarded as nearly 
independent of the system under study. 
A good general approximation to $W(\gamma_{12})$ can then be 
obtained by applying such a scaling to the functional dependence 
extracted from a simple reference system which already contains
the fundamental relationship between localization 
and correlation. We therefore derive an approximate $W(\gamma_{12})$ 
taking its functional dependence from the exact result for the 
Hubbard dimer given by Eq.~(\ref{eq:W2ex}). In this way one obtains
\begin{equation}
W(\gamma_{12}) = E_{\rm HF} 
\left( 1 - \sqrt{1 - 
\frac{(\gamma_{12} - \gamma_{12}^\infty)^2}
     {(\gamma_{12}^0 - \gamma_{12}^\infty)^2 }
                } \right) \; ,
\label{eq:W}
\end{equation}
where $E_{\rm HF}$, $\gamma_{12}^0$ and $\gamma_{12}^\infty$ 
are system specific [see (i)--(iv) above].
In practice, $\gamma_{12}^\infty$ may be approximated by the
ferromagnetic fully-polarized $\gamma_{12}^{\rm FM}$ which is 
calculated, as $\gamma_{12}^0$, by integration of the single-particle 
spectrum. 

\begin{figure}
\includegraphics[scale=0.45]{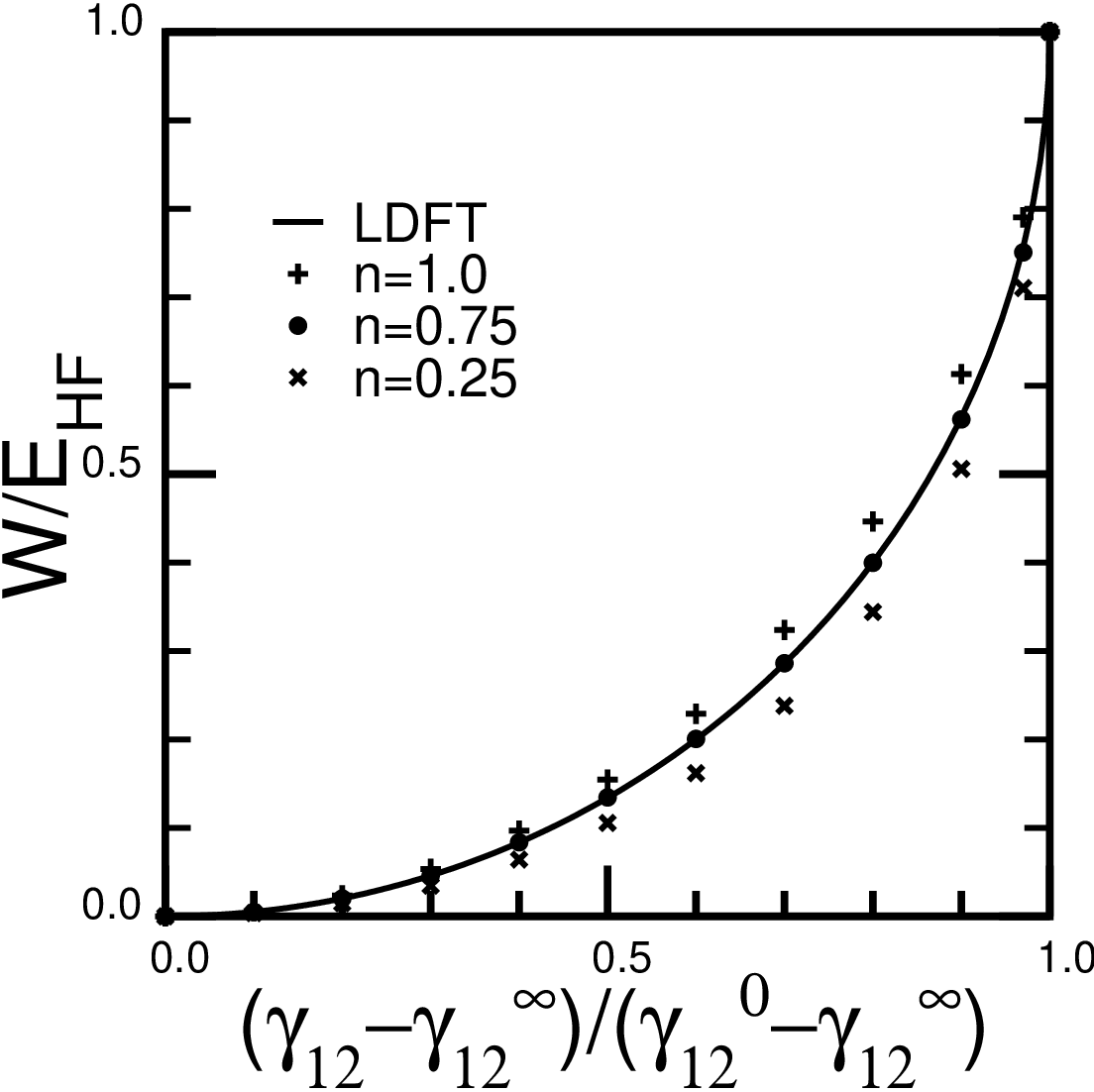}
\caption{\label{fig:Exc}
Interaction energy $W(\gamma_{12})$ of the one-dimensional 
(1D) Hubbard model 
as a function of the degree of electron delocalization 
$(\gamma_{12}   - \gamma_{12}^\infty) / 
(\gamma_{12}^0 - \gamma_{12}^\infty)$.
The symbols refer to exact results for different band fillings $n$
and the solid curve to Eq.~(\protect\ref{eq:W}).
        }
\end{figure}

Fig.~\ref{fig:Exc} compares Eq.~(\ref{eq:W}) with the exact 
$W_{ex}(\gamma_{12})$ of the 1D Hubbard chain which is derived from the 
Bethe-Ansatz solution.\cite{lieb-wu} One observes that the proposed 
approximation follows $W_{ex}(\gamma_{12})$ quite closely all along 
the crossover from weak correlations (large $W/U$ and $\gamma_{12}$) 
to strong correlations (small $W/U$ and $\gamma_{12}$). This is 
remarkable, taking into account the simplicity of Eq.~(\ref{eq:W}) 
and the strong band-filling dependence of $E_{\rm HF}$, $\gamma_{12}^0$, and 
$\gamma_{12}^\infty$. The quantitative discrepancies between 
Eq.~(\ref{eq:W}) and $W_{ex}(\gamma_{12})$ remain small in the 
complete domain of representability of $\gamma$ and for all band 
fillings: $|W - W_{ex}|/E_{\rm HF} \le 0.063$ for all $\gamma_{12}$ and $n$. 
Consequently, a good general performance of the method can be expected 
already at this stage. In the following section several applications 
of LDFT are discussed by using Eq.~(\ref{eq:W}) as approximation 
to the interaction-energy functional.

\section{\label{sec:res}Results and discussion}

\begin{figure}
\includegraphics[scale=0.5]{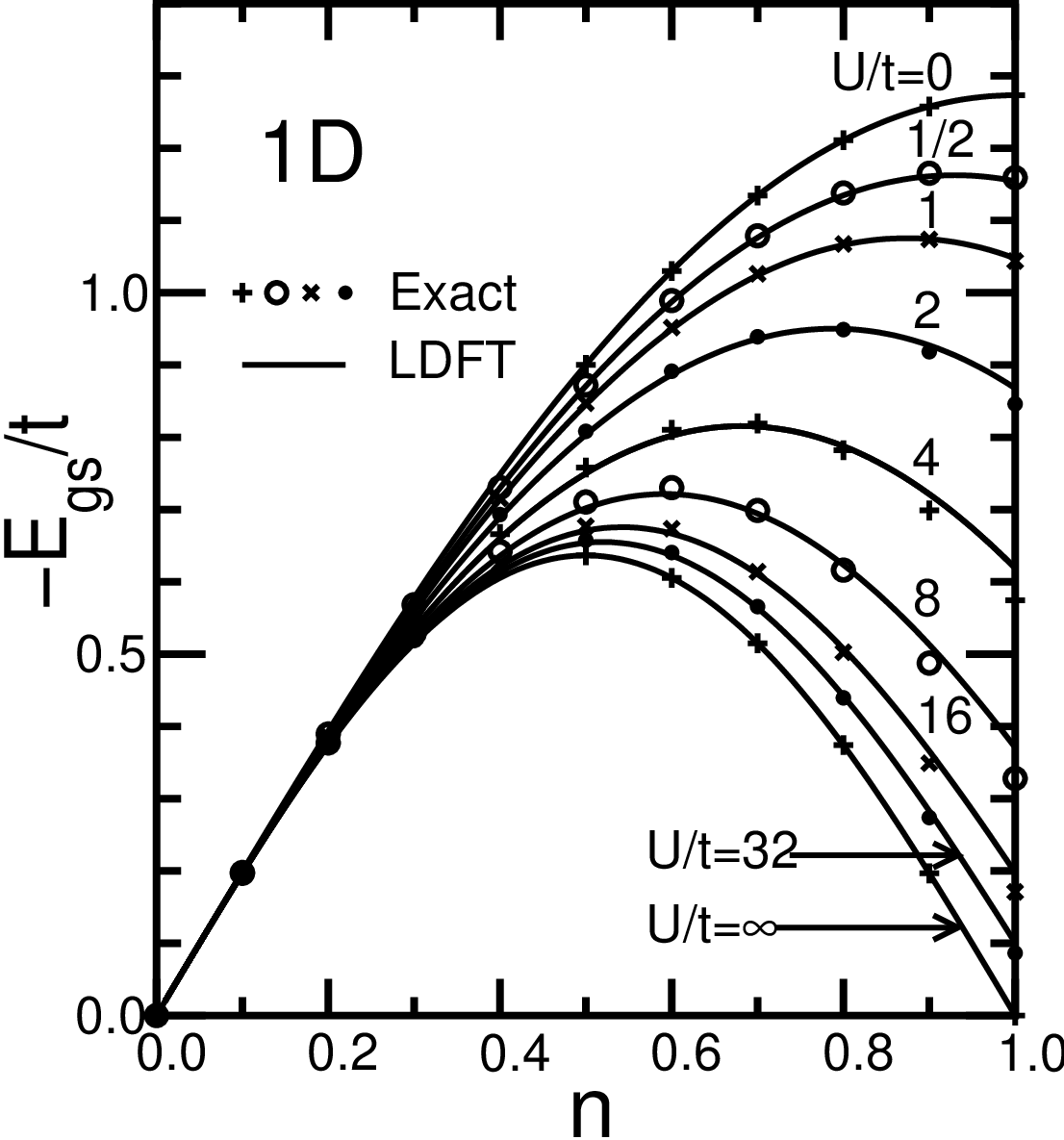}
\caption{\label{fig:E1D}
Ground-state energy $E_{gs}$ of the 1D Hubbard model as a function of 
band filling $n$ for different Coulomb repulsions $U/t$. The solid 
curves refer to the present lattice density-functional theory (LDFT) 
and the symbols to the Bethe-Ansatz exact solution.\protect\cite{lieb-wu}  
         }
\end{figure}

\begin{figure}
\includegraphics[scale=0.55]{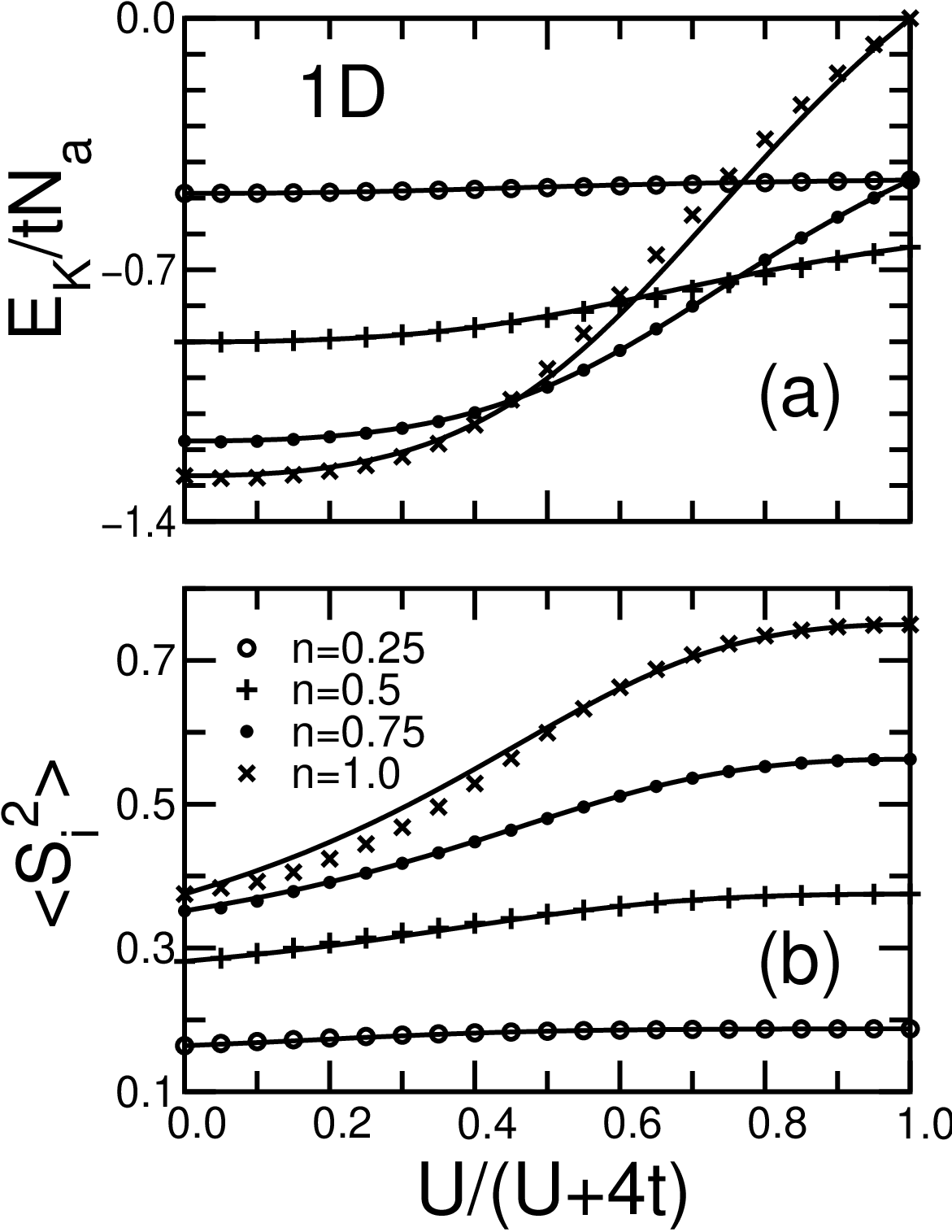}
\caption{\label{fig:ekin1D}
(a) Kinetic energy $E_K$ and (b) local magnetic moments 
$S_i^2 = 3 \langle (\hat{n}_{i \uparrow}- \hat{n}_{i \downarrow})^2 \rangle$ 
of the 1D Hubbard model as a function of Coulomb repulsion $U/t$ for 
different band fillings $n$ as indicated in the inset of subfigure (b). 
The solid curves correspond to the present LDFT and the symbols to 
the Bethe-Ansatz exact solution.\protect\cite{lieb-wu}  
         }
\end{figure}

In Fig.~\ref{fig:E1D} the ground-state energy $E_{gs}$ of the 1D 
Hubbard model is given as a function of band filling $n$ for different 
values of Coulomb repulsion $U/t$. Comparison between
LDFT and the Bethe-Ansatz exact solution shows a very good agreement.
It is interesting to observe that the accuracy of the calculated $E_{gs}$
is not the result of a strong compensation of errors since a similar 
accuracy is achieved for the kinetic and Coulomb energies 
separately. Indeed, as shown in Fig.~\ref{fig:ekin1D}, both local moments 
$S_i^2 = 3 \langle (\hat{n}_{i \uparrow}- \hat{n}_{i \downarrow})^2 \rangle$ 
and kinetic-energy renormalizations are also very well reproduced
as a function of $U/t$. Moreover, notice that no artificial symmetry 
breaking is required in order to describe correctly the 
correlation-induced localization, as it is often 
the case in other approaches (e.g., antiferromagnetic spin-density 
wave for $n=1$). For $n \le 0.8$, the LDFT results are almost 
indistinguishable from the exact ones. Even the largest quantitative 
discrepancies, found for $n=1$ and intermediate $U/t$, 
are acceptably small (e.g., $|E_{gs}-E_{gs}^{ex}|/t=0.044$ for $U/t=4$). 
For $U \gg t$ and $n=1$ we obtain $E_{gs} \simeq -\alpha t^2/U$ with 
$\alpha\simeq 3.24$ while the exact result is $\alpha=4\ln 2\simeq 2.77$. 
The error in the coefficient $\alpha$ can be corrected by including in 
Eq.~(\ref{eq:W}) a 4th-order term in $g_{12}$ which provides in addition 
with a systematic improvement for all values of the interaction strength   
($|E_{gs}^{ex}-E_{gs}|/|E_{gs}^{ex}| < 0.02$ for all $U/t$).\cite{tobe} 

\begin{figure}
\includegraphics[scale=0.55]{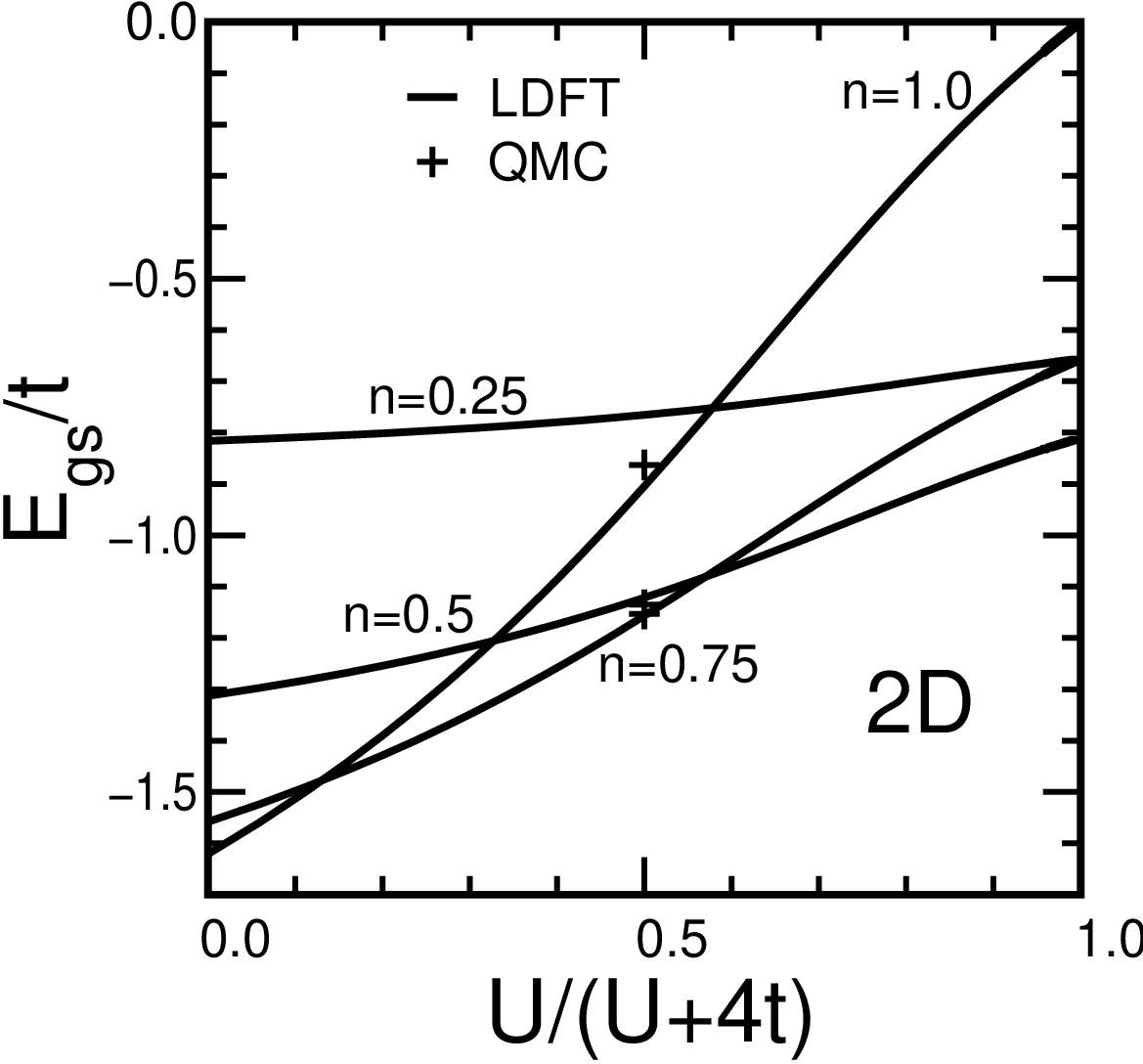}
\caption{\label{fig:E2D}
Ground-state energy $E_{gs}$ of the Hubbard model on the 2D square lattice 
as a function of Coulomb-repulsion strength $U/t$ and for different 
band fillings $n$. The solid curves refer to the present LDFT and the 
crosses to quantum Monte Carlo (QMC) calculations.\protect\cite{QMC-2D} 
        }
\end{figure}

\begin{figure}
\includegraphics[scale=0.55]{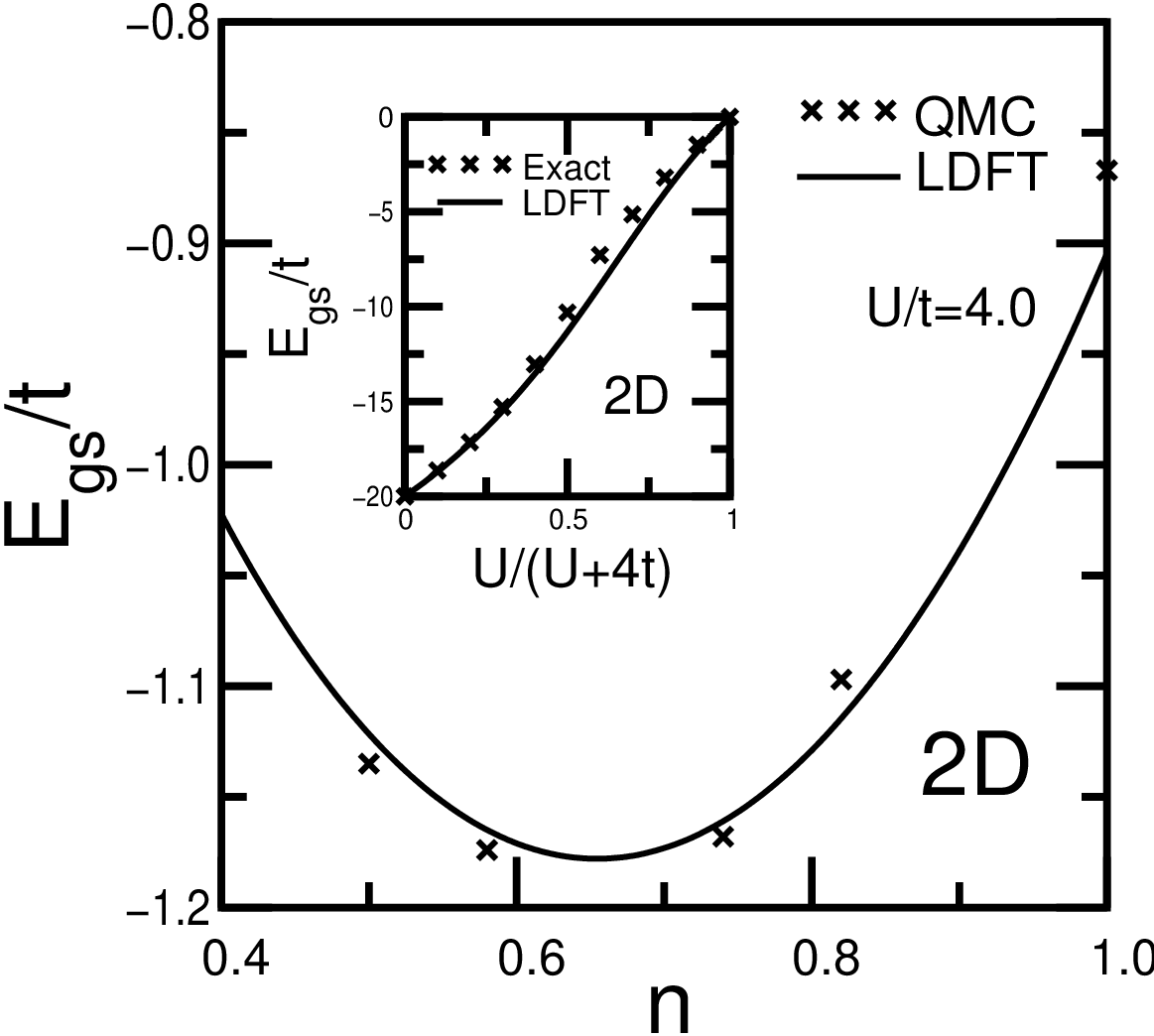}
\caption{\label{fig:Eclus2D}
Ground-state energy $E_{gs}$ of the Hubbard model on the 2D square 
lattice as a function of band filling $n$ for $U/t = 4$. The solid 
curve refers to the present lattice density-functional theory (LDFT) and 
the crosses to quantum Monte Carlo (QMC) calculations\protect\cite{QMC-2D}. 
In the inset figure LDFT is compared to exact Lanczos diagonalizations 
for a $N_a = 3 \times 4$ cluster of the 2D square lattice with periodic
boundary conditions. Results are here given as a function of $U/t$ at 
half-band filling ($n = N_e/N_a = 1$). 
        }
\end{figure}

Fig.~\ref{fig:E2D} shows $E_{gs}$ of the 2D square lattice as
a function of $U/t$ for representative band-fillings $n$. The LDFT 
results cover the complete range of model parameters involving 
essentially analytic calculations. As shown in Fig.~\ref{fig:Eclus2D}, 
good agreement is obtained with far more demanding ground-state 
quantum Monte Carlo (QMC) studies\cite{QMC-2D} for $U/t=4$. 
The reliability of LDFT in 2D systems is confirmed by comparison
with exact Lanczos diagonalizations on small clusters of the square and
triangular lattices. In the inset of Fig.~\ref{fig:Eclus2D} we
consider for example a $N_a= 3 \times 4$ cluster of the square lattice 
with periodic boundary conditions and $N_e = N_a$. Like in the 1D case, 
the overall performance is very good, with the largest quantitative
discrepancies being observed for intermediate values of $U/t$. 
For instance, for $U/t=1$ one obtains 
$|E_{gs}-E_{gs}^{ex}|/|E_{gs}^{ex}|= 4.4 \times 10^{-3}$, and for $U/t=4$ 
$|E_{gs}-E_{gs}^{ex}|/|E_{gs}^{ex}|= 9.8 \times 10^{-2}$. 
Results with similar precision are found for a the triangular 
2D lattice. In this case, using also 
a  $N_a= N_e = 3 \times 4$ cluster with periodic boundary conditions, 
we find $|E_{gs}-E_{gs}^{ex}|/|E_{gs}^{ex}|= 1.7 \times 10^{-4}$ 
for $U/t=1$, and $|E_{gs}-E_{gs}^{ex}|/|E_{gs}^{ex}|= 6.6 \times 10^{-2}$ 
for $U/t=4$. For both lattice structures $|E_{gs}-E_{gs}^{ex}|$ 
decreases quite rapidly away from half-band filling as in the 1D chain 
(see Fig.~\ref{fig:E1D}). LDFT, combined with Eq.~(\ref{eq:W}) 
for $W(\gamma_{12})$, provides a correct description of electron
correlations in different dimensions and lattice structures.

\begin{figure}
\includegraphics[scale=0.55]{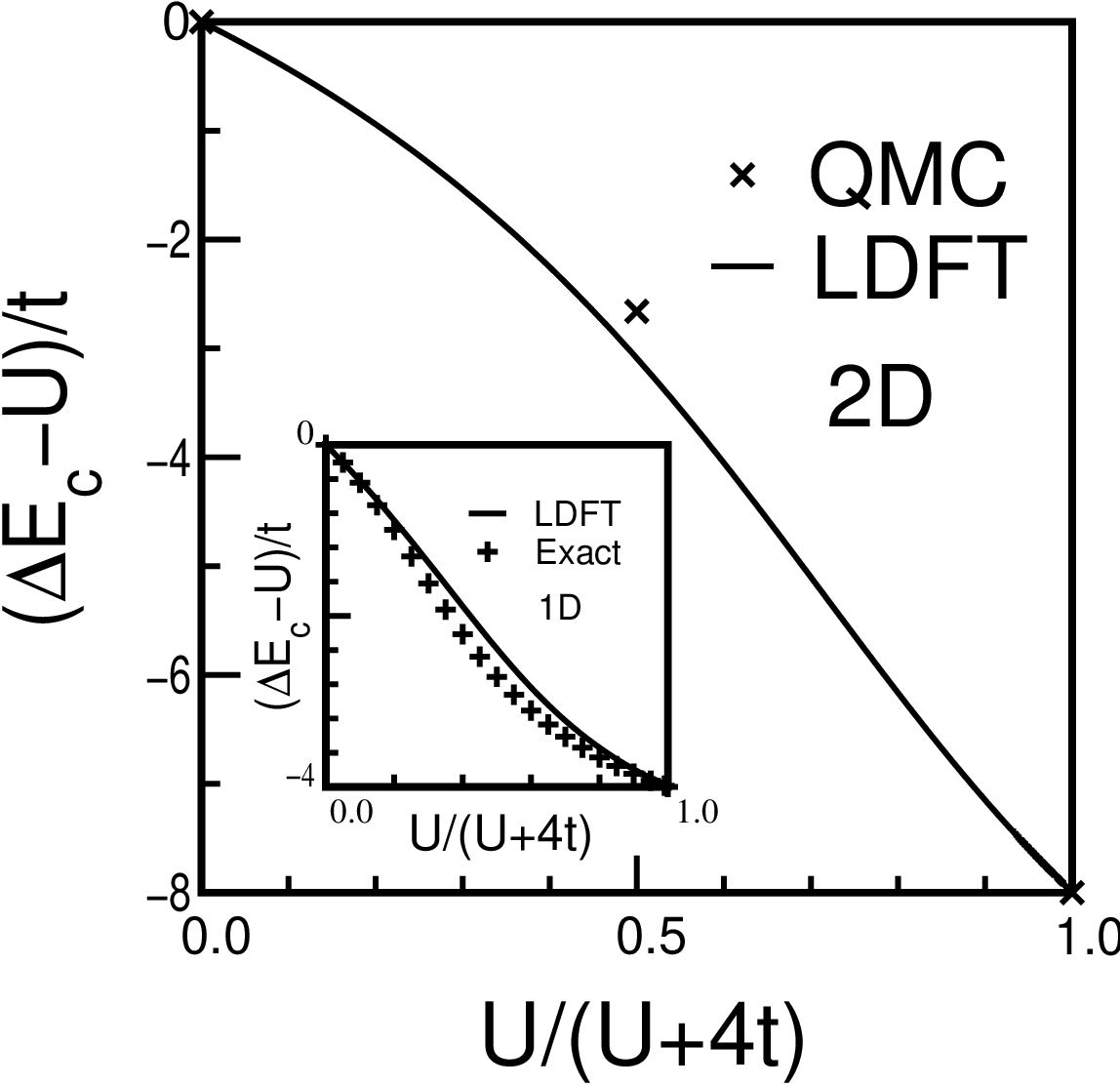}
\caption{\label{fig:gaps}
Charge excitation gap $\Delta E_c$ of the 2D Hubbard model (square 
lattice, $n=1$). In the inset results are given for the 1D chain. 
The solid curves refer to the present LDFT and the crosses to QMC 
calculations (2D, $U/t=4$) or to exact Bethe-Ansatz results 
(1D).\protect\cite{QMC-2D,lieb-wu} 
        }
\end{figure}

The charge-excitation or band gap 
\begin{equation}
\label{eq:gap}
\Delta E_c =  E_{gs}(N_{e}+1) + E_{gs}(N_{e}-1) - 2E_{gs}(N_{e})
\end{equation}
is a property of considerable interest in strongly correlated systems 
which measures the insulating or metallic character of the electronic 
spectrum as a function of $U/t$ and $n$. It can be directly related to 
the discontinuity in the derivative of the kinetic and correlation 
energies per site with respect to electron density $n$.\cite{foot:gap} 
Therefore, the determination of $\Delta E_c$ constitutes a much more 
serious challenge than the calculation of $E_{gs}$, particularly 
in the framework of a density-functional formalism. At half-band filling,
$\Delta E_c = 0$ in the uncorrelated limit ($U/t = 0$) and it increases 
with increasing $U/t$. For $U/t \to \infty$, 
$\Delta E_c \to U + E_b$ where $E_b$ is the energy of the bottom of 
the single-particle band ($E_b = -4t$ for a 1D chain and 
$E_b = -8t$ for the 2D square lattice). 
Fig.~\ref{fig:gaps} presents LDFT results for $\Delta E_c$ 
in 1D and 2D Hubbard models ($n=1$). Comparison with 
the Bethe-Ansatz results\cite{lieb-wu} and with available 
QMC calculations\cite{QMC-2D} shows a good overall agreement.
However, a more detailed analysis reveals that in the 1D case the  
gap is significantly overestimated for $U/t\ll 1$. Here we obtain 
$\Delta E_{\rm c} \propto (U/t)^2$, while the exact solution 
shows that for the infinite chain $\Delta E_{\rm c}$ increases much 
more slowly, namely, exponentially in $-t/U$. This discrepancy 
reflects the difficulty to describe long range-effects using an 
interaction-energy which functional dependence is derived from the 
dimer. Thus, it is possible that a similar overestimation of the gap 
at small $U/t$ may also affect our results on 2D lattices.
For larger $U/t$ the accuracy improves rapidly as electron localization 
starts to set in, and the relative error in $\Delta E_{\rm c}$ vanishes.
Therefore, the development of a Mott insulator with increasing $U/t$
is described correctly.

Finally, we would like to comment briefly on a few other applications: 
(i) Dimerized chains with hoppings $t \pm \delta t$ have been investigated
by allowing for alternations of $\gamma_{12}$ in Eq.~(\ref{eq:W}). 
One observes that the precision of the results improves systematically 
with increasing dimerization. For example, for $N_a=N_e=12$ and $U/t=4$,
we find $|E_{gs}-E_{gs}^{ex}|/|E_{gs}^{ex}| = 3.3 \times 10^{-2}$, 
$1.4 \times 10^{-2}$, and $2.6 \times 10^{-3}$ for
$ \delta t/t=0$, $1/2$, and $3/4$, respectively. 
The non-dimerized case, shown in detail in Fig.~\ref{fig:E1D}, 
is in fact the most difficult one, since for a collection of 
dimers ($\delta t=t$) the exact $W$ is recovered [Eq.~(\ref{eq:W2ex})]. 
(ii) Three dimensional (3D) lattices are a further interesting direction 
for future developments. Indeed, encouraging results have been obtained 
for the simple cubic lattice at half-band filling. LDFT with
Eq.~(\ref{eq:W}) for $W$ yields $E_{gs}/t=1.21$, $0.81$, and $0.59$ 
for $U/t=4$, $8$ and $12$, respectively, in good agreement with 
corresponding quantum Monte Carlo results,\cite{QMC-3D}
namely, $E_{gs}^{\rm QMC}/t = 1.27$, $0.78$, and $0.57$.
(iii) An accurate approximation to $W(\gamma_{12})$ has been also derived for
the  attractive (negative $U$) Hubbard model in an analogous way as 
for $U>0$. For a 1D ring with $N_a=N_e=12$ we find 
$|E_{gs}-E_{gs}^{ex}|/|E_{gs}^{ex}|=1.2\times 10^{-3}$, 
$7.5\times 10^{-3}$, and $1.5\times 10^{-4}$ for $|U|/t=1$, $4$,
and $64$, respectively. These results show that LDFT describes 
electronic correlations correctly also when intra-atomic pairing 
is favored. Systematic investigations along these lines are currently 
in progress and will be published elsewhere.\cite{tobe}

\section{\label{sec:concl}Conclusion}

A new density-functional approach to lattice-fermion models has been 
developed that is by all means independent of the homogeneous electron 
gas. A simple approximation to the interaction-energy functional is 
derived for the Hubbard model, which provides with a unified description 
of correlations in all interaction regimes from weak to strong coupling. 
Results for the ground-state energy and charge-excitation gap of
1D and 2D systems demonstrate the ability of lattice density
functional theory to describe quantitatively the subtle competition 
between kinetic charge fluctuations and correlation-induced localization. 
The scope of DFT is thereby extended to the limit of strong electron 
correlations.

Several interesting directions open up with potential implications in 
various related areas. For example, one may explore more general 
approximations to $W[\gamma]$ and one may apply the present approach to 
richer physical situations such as low-symmetry systems, disorder, 
magnetic impurities, or multiband Hamiltonians. These developments should 
be relevant to the study of lattice fermion models and also in view of 
a DFT description of strong correlations from first-principles.

\begin{acknowledgements}
One of the authors (RLS) acknowledges financial support from CONACyT (Mexico)
through the project W-8001 (Millennium initiative).
Computer resources were provided by IDRIS (CNRS, France).
\end{acknowledgements}

%

%
%


\begin{thebibliography}{}
\bibitem{foot_local}
A formulation of DFT on a lattice only in terms of the diagonal 
$\gamma_{ii}$ is possible only if one restricts oneself to models 
with constant $t_{ij}$ for $i\not= j$. In this case the functional 
$W[\gamma_{ii}]$ depends on the values of $t_{ij}$ for $i\not= j$ 
and in particular on $U/t$ (see Ref.~\protect\cite{gunn}).
\bibitem{foot_gen}
The LDFT formalism can be readily extended to arbitrary interactions
$H_I = (1/2) \sum V_{ijkl} 
\hat c_{i \sigma }^{\dagger} \hat c_{k   \sigma '}^{\dagger} 
\hat c_{l \sigma '} \hat c_{j \sigma}$, by replacing 
$U \sum_i \hat n_{i\uparrow} \hat n_{i\downarrow}$ by $H_I$ in 
Eq.~(\ref{eq:Wlevy}). However, note that such a general functional 
$W[\gamma]$ depends crucially on $V_{ijkl}$. See also 
Ref.~\protect\cite{prbxcfun}.
\end{thebibliography}
\end{document}